%% file: etacpaper.tex
\documentclass[12pt,a4paper,dvips]{article}
\usepackage{a4p}
\usepackage{cite,mcite}
\usepackage{graphicx}
\usepackage{physics }
\usepackage{l3_title,ifthen,Lep}
\def\chiczero{\chi_{{\rm c}0}}
\def\chicone{\chi_{{\rm c}1}}
\def\chictwo{\chi_{{\rm c}2}}
\def\jpsi{{\rm J}}
\def\gaga{\gamma \gamma}
\def\Ggg{\ensuremath{\Gamma_{\gamma\gamma}}}%
\def\qsq{Q^{2}}
\def\etac{\eta_{c}}
\def\pipi{\pi^{+}\pi^{-}}
\def\ks{{\rm K}_{s}^{0}}
\def\k0kpi{{\rm K}^{0}_{s} {\rm K}^{\pm} \pi^{\mp}}
\def\kk{{\rm K}^{+}{\rm K}^{-}}
\def\rhrh{{\rho}^{+}{\rho}^{-}}
\def\sumpt{(\Sigma \vec{p}_{\rm t})^2}
\def\beq{\begin{equation}}
\def\eeq{\end{equation}}
\journalname{Phys. Lett.}
\date{May 21, 1999}
\preprint{99-072}
\newlength{\capindent}
\newlength{\capwidth}
\newlength{\figwidth}
\newcommand{\icaption}[2][!*!,!]{\hspace*{\capindent}%
  \begin{minipage}{\capwidth}
    \ifthenelse{\equal{#1}{!*!,!}}%
      {\caption{#2}}%
      {\caption[#1]{#2}}
  \end{minipage}}
\begin{document}
 
\begin{titlepage}

\title {\boldmath Formation of the $\etac$ in Two-Photon Collisions at LEP}
\author{The L3 Collaboration}

%
%
\begin{abstract}

The two-photon width $\Ggg$ of the $\etac$ meson has been
measured with the L3 detector at LEP. The $\etac$ is studied in the 
decay modes $\pipi\pipi$, $\pipi\kk$, $\k0kpi$, $\kk\pi^{0}$, $\pipi\eta$, 
$\pipi\eta'$, and $\rhrh$ using an integrated luminosity of 140 pb$^{-1}$
at $\sqrt{s} \simeq  91$ GeV and of 52 pb$^{-1}$ at $\sqrt{s} \simeq 183$ GeV.
The result is
$\Ggg(\etac) = 6.9 \pm 1.7 \stat \pm 0.8 \sys \pm 2.0$(BR) \keV.
The $\qsq$ dependence of the $\etac$ cross section is studied for 
$\! \qsq<\! 9 \GeV^{2}$. It is found to be better described by a Vector 
Meson Dominance model form factor with a $\jpsi$-pole than 
with a $\rho$-pole.
In addition, a signal of 29$\pm$11 events is observed at the $\chiczero$ mass.
Upper limits for the two-photon widths of the $\chiczero$, $\chictwo$, 
and $\etac'$ are also given.

\end{abstract}
%
%
\submitted

\end{titlepage}
%
%
\section*{Introduction}

The study of resonance formation via
two-photon interactions in $\ee$ colliders provides valuable
information on the quark substructure of the meson. 
The cross section for two-photon resonance formation
$\ee\ra\ee\gamma^{*}\gamma^{*}\ra \ee R$, where
$R$ is a  C=+1 meson, is given by \cite{budnev}
\begin{equation}
  \sigma({\rm e^{+} e^{-}} \ra \ee R) = 
  \int \sigma_{ \gaga \ra R} ~d  L_{\gaga} (W^{2}),
  \label{eq:sigmatot}
\end{equation}
with the Breit-Wigner cross section 
\begin{equation}
  \sigma_{ \gaga \ra R}(W^{2},q_{1}^{2},q_{2}^{2}) = 
  8 \pi  (2 J_{R}+1) \Ggg(R) F^{2}(q_{1}^{2},q_{2}^{2}) \cdot   
  \frac{  \Gamma_{R}}{ (W^{2}-M_{R}^{2})^{2} + 
           M_{R}^{2} \Gamma_{R}^{2}}.
\label{eq:sigmagg}
\end{equation} 
Here, $W$ is 
the two-photon centre of mass energy, $q_{1}^{2}$ and $q_{2}^{2}$ 
are the squares of the virtual photon four-momenta, and 
$ L_{\gamma \gamma} (W^{2})$ is the two-photon luminosity function.
The resonance $R$ is characterised by the mass $M_{R}$, total spin $J_{R}$,  
total width $\Gamma_{R}$ and the
two-photon partial width $\Ggg(R)$. The two-photon width and the transition
form factor $F^{2}(q_{1}^{2},q_{2}^{2})$ are the two parameters to be measured.

In two-photon collisions, the scattered electron and positron 
emerge with a small energy loss and an almost unmodified 
direction after having radiated one photon each. 
They therefore usually go undetected along the beam direction, 
allowing the photons to be considered
as `quasi-real', with $q^{2}\simeq 0$.
At $q_{1}^{2}\simeq q_{2}^{2}\simeq 0$, the form factor 
$F^{2}(q_{1}^{2},q_{2}^{2})$ is normalised to unity, leaving the 
two-photon width as the only unknown parameter, linearly proportional 
to the total cross section:
\begin{equation}   
  \sigma(\ee \ra \ee R) = {\kappa} \; \Ggg(R).
\label{eq:kappa}
\end{equation}
 
If the scattered electron or positron is detected at small angles, 
the event is said to be {\it tagged}: one photon has $Q^{2}=-q^{2}$
while the other is quasi-real with $q^{2}\simeq 0$.
Using tagged events, the transition
form factor can be measured. It is parametrised
in the Vector Meson Dominance (VMD) model by a pole form 
\begin{equation}
  F (Q^2) =  \frac{1}{1+Q^2 / \Lambda^2}   
  \;\;\;\;  \mathrm{with} \;\;\;\;
\Lambda^2 = M_{V} ^2 , \;\;\;\; V= \rho,\omega,\phi,\jpsi ...
\label{eq:ff}
\end{equation}

In the case of charmonium (\ccbar) mesons, the $J^{PC}$ states
accessible in quasi-real two-photon reactions are 0$^{-+}$, 0$^{++}$ 
and 2$^{++}$,
corresponding to $\etac(2979)$, 
$\chiczero(3417)$ and $\chictwo(3556)$.
The formation of the $\chicone(3510)$, a 1$^{++}$ state, is forbidden for 
two real photons according to the Landau-Yang theorem \cite{yanglandau}.
The radial recurrence $\etac'(3594)$ has been observed by the Crystal Ball 
experiment \cite{etacprime} in the process $\psi'\ra\gamma\etac'$, but 
not in two-photon collisions \cite{delphi_etacp}.

In this paper we report on the study of the formation of the $\etac$.
The two-photon width and the $\qsq$ dependence of the cross section 
are determined.
Theoretical calculations, based on the assumption 
that the two heavy charm quarks are bound together 
by a QCD potential, predict 3 - 9 \keV\ for the $\etac$ two-photon width
 \cite{theoryetac}.
The $Q^2$ dependence of the cross section has also been calculated 
\cite{feldmann,schuler}. 
It contains information about the quark momentum
distribution inside the meson, and
its shape is predicted to be well described by a 
VMD model form factor with $M_{V} = M_{\jpsi}$.
The $\qsq$ dependence of the $\etac$ cross section has not been measured 
up to now.

We report results from the data obtained with the
L3 detector at centre of mass energies $\rts \simeq$ 91 \GeV,
with a total
integrated luminosity of 140.2 \pb, and at 
$\rts \simeq$ 183 \GeV, with an integrated luminosity of 52.4 \pb. 
This data sample includes the 30 pb$^{-1}$
which we used for our previous measurement of $\Ggg(\etac)$ \cite{l3etac}. 
Since the $\etac$ resonance does not have a dominant decay mode, nine
different decay modes with branching ratios ranging from 0.5\% to 2\% 
have been analysed.
They are listed in  Table \ref{table:result}
together with the branching ratio of each channel, derived from Reference 
\cite{PDG98}.

%
%
\section*{L3 detector and  Monte Carlo}

A detailed description of the L3 detector can be found in 
Reference \cite{l3detector}.
The analysis described in this paper is mainly based on 
the central tracking system and   
the high resolution electromagnetic calorimeter.

Particles scattered at small angles are measured
by the luminosity monitors (LUMI), 
covering a polar angle range  26 ${\rm mrad}<\theta<65$ mrad
on each side of the detector.
For data at $\rts\simeq 183 \GeV$, particles can also be detected
by the very small angle tagger (VSAT) covering a polar angle range between
5 ${\rm mrad}<\theta<8$ mrad for an azimuthal angle range of
-0.8 rad$<\! \phi\! <$0.8 rad or 
$\pi$-0.8 rad$<\! \phi\! <\! \pi$+0.8 rad \cite{thesistasja}.

The two-photon events are collected 
predominantly by a track trigger \cite{trigger} which requires
at least two charged particles with transverse momentum $p_{\rm t} >$ 150 \MeV,
back to back, in the plane transverse to the beam, 
within $\pm 41^{\circ}$ for data at $\rts\simeq 91 \GeV$, 
and within $\pm 60^{\circ}$ for data at $\rts\simeq 183 \GeV$.
In addition, there is a trigger for electron tags, which requires 
an energy deposit of at least 70\% of the beam energy in the LUMI, 
in coincidence with at least one charged track in the central tracker.

In order to compute the acceptance and
efficiency of the detector 
we have used the PC Monte Carlo \cite{thesisfl}, based on the formalism of
Budnev \etal \cite{budnev}.
The $Q^{2}$ dependence of the cross section is
taken into account using a VMD model form factor.

The particles produced in the reaction are followed through
the different L3 subdetectors with the GEANT \cite{mygeant} simulation
program  and the events are 
reconstructed and analysed in the same way as real data.

%
%
\section*{Selection criteria}

The $\etac$ decays,
listed in Table \ref{table:result}, result in
final states with either two or four charged 
tracks with charge balance, accompanied
by zero, one, two or four photons.

A track must have at least 18 out of a maximum of 62 hits in the 
central tracker. 
The distance of closest 
approach to the beam line in the transverse plane is required to be 
less than 3 mm, except for tracks associated with a $\ks$ decay. 
In order to remove electrons,
we require for 
tracks with a momentum $p$ larger than 0.8 \GeV\ that $E/p<0.9$, where 
$E$ is the associated energy in the electromagnetic calorimeter.

A photon candidate
is an electromagnetic cluster 
separated from all tracks by at least 100 mrad in $\phi$ and 140 mrad 
in $\theta$. 
In the search for $\pi^{0}$ candidates, the two photons must both have
an energy of at least 50 \MeV. 
To reduce the combinatorial background under the 
$\pi^{0}$ signal, 
the angle $\psi_{\gaga}$ 
between the photons of the $\pi^{0}$ candidate must satisfy  
$\cos\psi_{\gaga}>0.6$. The effective
mass of the two photons must be within 
20 \MeV\ of the $\pi^0$ mass if both photons are in the barrel 
part of the electromagnetic calorimeter ($42^{\circ}<\theta<138^{\circ}$), 
and within 30 \MeV\ otherwise. The $\pi^{0}$ signal is shown in  
Figure \ref{fig:mcuts}a and \ref{fig:mcuts}b.
The two photons of an $\eta$ candidate 
must have an energy of at least 100 \MeV,
and at least one of them must be in the barrel.
In addition, it is required that 
$\cos\psi_{\gaga}\! > \! -0.7$, 
and that the $\gaga$ effective mass is within 45 \MeV\ of the $\eta$ mass, 
see Figure \ref{fig:mcuts}c.

If a cluster with at least 70\% of the beam energy is found in the 
LUMI calorimeter or a cluster with more than 50\% of the beam energy is found
in the VSAT, the event is classified as tagged.

To ensure that no final state particle of the resonance decay has escaped 
detection, the squared vectorial sum of the 
transverse momenta of all detected 
particles, $\sumpt$, is required to be smaller 
than 0.1 \GeV$^2$. If a tag is found, it is also included in this sum.
However, if the $\sumpt$ excluding the tag is smaller than the $\sumpt$
with the tag included, the event is considered as untagged.
Additionally, for events with one or more photons in the final state,
the $\sumpt$ excluding the photon(s) should be larger than 0.005 \GeV$^2$.

Additional cuts for each decay mode are listed below. The masses of the 
intermediate states which are used in the cuts are taken from the observed 
central values of the peaks.

\medskip

{\noindent \underline {$\etac \ra  \pipi\pipi$ or $\pipi\kk$}}\\
These events leave four tracks and no photons 
in the detector.
Using the $dE/dx$ measurement in the central tracking system, a 
$\chi^{2}$ to be a pion or kaon can be calculated for each track, 
and thus a combined probability that the four tracks are $\pipi\pipi$,
$\kk\pipi$, or $\kk\kk$ can be derived.
For a $\pipi\pipi$ event, the $\pipi\pipi$ probability divided by the sum of
$\pipi\pipi$, $\kk\pipi$, and $\kk\kk$ probabilities must be larger than 55\%.
For a $\kk\pipi$ event, the $\kk\pipi$ probability divided by the sum of
$\pipi\pipi$, $\kk\pipi$, and $\kk\kk$ probabilities must be larger than 55\%.
Events that do not fall in either category are rejected.
The $dE/dx$ measurement has a good $\pi/K$ separating
power for tracks with a momentum $p<0.5$ \GeV. As the momenta of the
decay particles 
of the $\etac$ extend to 2 \GeV, there exists some misidentification, 
which is estimated using Monte Carlo.
Out of all $\etac\ra\pipi\pipi$ Monte Carlo events that 
are selected, 21\% are identified wrongly as $\kk\pipi$, and of
all $\etac\ra\kk\pipi$ Monte Carlo events that are selected, 
8\% are wrongly identified as $\pipi\pipi$.

\medskip

{\noindent\underline {$\etac \ra \k0kpi$}}\\
For this final state the
geometrical reconstruction of the secondary vertex
$\ks \ra \pipi$ requires
two oppositely charged tracks, each with a radial distance from the 
interaction point in the transverse plane
greater than 1 mm.
The two tracks must form a secondary vertex
more than 3 mm away from the interaction point. The angle between 
the tracks must be smaller than 2.5 rad.
If the angle between the $\ks$ line
of flight and the sum of the two track momenta is greater than 60 mrad, the
$\ks$ candidate is rejected. Finally, the invariant mass of the 
two pions must be within 30 \MeV\ of the $\ks$ mass, 
see Figure \ref{fig:mcuts}d. 

\medskip

{\noindent\underline {$\etac \ra \pipi\eta, \, \eta \ra \gaga$} \\
For this decay mode 
two photons, reconstructing an $\eta$, and two tracks 
must be observed in the detector.

\medskip
{\noindent\underline {$\etac \ra \kk\pi^0$}}\\
For this decay mode 
two photons, reconstructing a $\pi^0$, and two tracks must be observed 
in the detector. Since there exists 
no background from $\etac\ra\pipi\pi^{0}$, only a loose cut on the $dE/dx$
is used in order to keep the efficiency high: 
the $\kk$ probability divided by
the sum of the $\kk$ and $\pipi$ probabilities must be larger than 30\%.

\medskip

{\noindent\underline {$\etac \ra \rhrh, \, \rho^{\pm} \ra \pi^{\pm}\pi^{0}$}} \\
Four photons are combined in order to find two $\pi^0$ candidates, 
which are combined 
with the two tracks to form two charged $\rho$ mesons.
The event is accepted if both combinations satisfy
$|M(\pi^{\pm}\pi^{0}) - M(\rho^{\pm})|<0.25 \GeV$.

\medskip

{\noindent\underline {$\etac \ra \pipi\eta, \, \eta \ra \pipi\pi^{0}$}}\\
For this decay mode the cut on the angle between the two photons 
is widened to $\cos\psi_{\gaga}>0.4$.
Two out of the four observed tracks must combine with the 
$\pi^0$ candidate to form an $\eta$, with
$|M(\pipi\pi^0)-M(\eta)|<40 \MeV$.

\medskip

{\noindent\underline {$\etac \ra  \pipi \eta', \, \eta' \ra \pipi\eta, 
\, \eta \ra \gaga$}}\\
For this decay mode two photons, reconstructing an $\eta$, and four tracks 
must be observed in the detector.
The $\eta$ candidate must combine with two of the tracks to form
an $\eta'$, with 
$|M(\pipi\eta)-M(\eta')|<50 \MeV$.
  
\medskip

{\noindent\underline {$\etac \ra  \pipi\eta', \, \eta' \ra \rho^{0}\gamma, 
\, \rho^{0} \ra \pipi$}}\\
To select this decay mode,  
$\rho^{0}$ candidates, formed  by a pair of tracks with
$0.60 \GeV < M(\pi^{+}\pi^{-})<  0.92 \GeV$, are combined with a photon, with
energy greater than 100 MeV, to form an $\eta'$, with 
$|M(\rho^{0}\gamma)-M(\eta')|<60 \MeV$ and 
$\cos\psi_{\rho^{0}\gamma}>-0.3$.  

\medskip

The selection efficiencies for the analysed
decay modes are given in Table \ref{table:result}, for both data samples
at $\rts\simeq 91 \GeV$ and at $\rts\simeq 183 \GeV$.

The mass distribution of the selected untagged events is
presented in Figure \ref{fig:m1etac}.
A fit of this spectrum in the mass range $2.4 \GeV < M <3.2 \GeV$, 
with an exponential background plus a 
Gaussian for the signal, gives for the peak position 
$M(\etac$)= 2.974 $\pm$ 0.018 GeV, in excellent
agreement with the world average 2.979 $\pm$ 0.002 GeV \cite{PDG98}.
The width of the Gaussian is 57$\pm$16 \MeV, consistent with
the reconstructed width in the Monte Carlo of 70 \MeV.
The area of the peak corresponds to a total of 
93$\pm$33 $\etac$ events.

The $\chiczero(3417)$ and $\chictwo(3556)$ 
are also known to decay into $\pipi\pipi$ 
and $\kk\pipi$. Their branching ratios to the other analysed decay modes 
are unknown. 
Therefore the mass spectrum has been divided into two parts in 
Figure \ref{fig:m2etac}: Figure \ref{fig:m2etac}a shows
the mass distribution
of the events with one or more photons or 
a $\ks$ in the final state, while 
Figure \ref{fig:m2etac}b shows the mass spectrum of the 
$\pipi\pipi$ and $\kk\pipi$ events.
However, in Figure \ref{fig:m2etac}b, no enhancements are visible around 
the $\chiczero$ and $\chictwo$ masses.
In Figure \ref{fig:m2etac}a, on the other hand, an 
enhancement is visible around 3.4 \GeV.
A fit to this enhancement in the mass range $3.2 \GeV <M<4 \GeV$ gives
for the peak position 3.400$\pm$0.019 \GeV, 
in agreement with the world average $\chiczero$
mass $M(\chiczero$)=3.417$\pm$0.003 \GeV, 
and for the width 58$\pm$18 \MeV, in agreement with the reconstructed 
width from the Monte Carlo of 70 MeV.
The area of the fitted Gaussian is 29$\pm$11 events.
A similar fit to Figure \ref{fig:m2etac}b yields only 7$\pm$6 events 
above background.


\section*{The two-photon width}

Since the $\etac$ signal has low statistics in each decay mode,
the two-photon width is obtained with a combined unbinned likelihood fit.
The fit takes into account the different weight of each decay mode
due to the different background levels, efficiencies, 
branching ratios, and the different integrated luminosities at 
$\rts\simeq 91 \GeV$ 
and at $\rts\simeq 183 \GeV$.
All mass spectra are fitted simultaneously with a sum of a normalised 
exponential background function $b_{i}(x)$ and a Gaussian distribution 
$g_{i}(x)$, where $i$ runs over the different decay modes and the two 
centre-of-mass energies considered:
\beq
f_{i}(x) = (1-p_{i}) b_{i}(x) + p_{i} g_{i}(x).
\eeq
Here, $p_{i}$ is the ratio of the number of signal events, $S_{i}$, over the 
total number of events in mass spectrum $i$. The exponential background, 
$b_{i}(x)$, is allowed to vary for each spectrum, but the number 
of signal events in the Gaussian for each spectrum is related to the 
two-photon width by 
\beq
\label{eq:S_i}
S_i = \epsilon_i \: \mathcal{L} \: {\rm BR}_i \: \kappa \: \Ggg.
\eeq
Here, $\epsilon_i$ is the total efficiency, $\mathcal{L}$ 
is the integrated luminosity,
and ${\rm BR}_{i}$ is the branching ratio for decay mode $i$.
The proportionality factor $\kappa$ is obtained from Monte Carlo, 
using Equation (\ref{eq:kappa}).
The events in $\chiczero$ signal region
(between 3.3 and 3.5 \GeV) are excluded from the fit.
For the $\kk\pipi$ and $\pipi\pipi$ mass spectra, the whole region $M>3.2 \GeV$
is excluded, where the $\chiczero$ signal, the $\chictwo$ signal, 
and small enhancements due to $\pi/K$ misidentification are expected.
The effect of this cut is taken into account in the systematic error.
The $\etac$ two-photon width obtained from the fit is $6.9\pm 1.7\stat 
\keV$. It corresponds to a total of 76$\pm$19 $\etac$ 
events, in agreement with the number of events obtained 
with the fit to the total mass spectrum shown in Figure \ref{fig:m1etac}.
The two-photon widths or upper limits 
for the individual decay modes are also given 
in Table \ref{table:result}. They have been obtained 
with an unbinned likelihood fit to the individual spectra using 
an exponential background and a Gaussian for the signal, with the 
Gaussian position and width fixed to the Monte Carlo values.

The systematic error related to the selection requirements includes 
contributions from the cut on the number of hits on a track (5\%), 
from the $dE/dx$ cuts (3\%), and from the $\pi^{0}$ selection (4\%).
Furthermore, the systematic errors 
take into account uncertainties on the trigger
efficiency (2\%), on the Monte Carlo statistics (2\%), 
on the background subtraction (3\%) and on
the uncertainty on the individual branching ratios (9\%).
They add up to 12\%, 
resulting in a systematic error on $\Ggg(\etac)$ 
of 0.8 \keV.
The uncertainty introduced by
the poor knowledge of the branching ratio
BR($\jpsi \ra \etac\gamma$) = (1.27$\pm$ 0.36) \% \cite{bretacg}, 
which is contained in all branching ratio uncertainties,
leads to an additional error on $\Ggg(\etac)$ of 2.0 \keV.

In Table \ref{table:comparison} our measurement is compared to previous 
measurements \cite{others,l3etac} of the $\etac$ two-photon width. 
It is found to be in good agreement. The measurement is also in good 
agreement with the theoretical predictions \cite{theoryetac}.

If we assume that the excess of events at 3.4 \GeV\ is due to
$\chiczero$ formation, the $\chiczero$ two-photon width can be obtained 
using a similar fit as for the $\etac$.  
However, only the $\pipi\pipi$ and $\kk\pipi$ events, for which the 
$\chiczero$ branching ratios are known, can be included in the fit.
Since the number of signal events is low for these decay modes,
it is only possible to set a 95\% C.L. upper limit 
$\Ggg(\chiczero)< 5.5\keV$.
This is consistent with the
upper limit by CLEO, 
$\Ggg(\chiczero)<6.2 \keV$, and the measurement by the 
Crystal Ball experiment,
$\Ggg(\chiczero)=4.0\pm 2.8 \keV$ \cite{chic0meas}.

Also the $\chictwo(3556)$ has known branching ratios 
into $\pipi\pipi$ and $\kk\pipi$.
No signal is observed in Figure \ref{fig:m2etac}b. 
We obtain for the $\chictwo$ two-photon width a 95\% C.L. 
upper limit of 1.4 \keV, consistent with our previous measurement
\cite{chic2paper}. 

We also observe no signal for the $\etac'(3594)$. 
Reference \cite{etacpbr} predicts the hadronic branching ratios 
of the $\etac$ and $\etac'$ to be about equal.
If we assume that the efficiencies and branching ratios of 
the $\etac'$ are the same as for the $\etac$ for the analysed 
final states, we obtain $\Ggg(\etac')<2.0 \keV$ at 95\% C.L.


\section*{The $\etac$ form factor}

The $\etac$ transition form factor, as defined in 
Equation (\ref{eq:sigmagg}), can be studied using tagged events. 
The mass spectrum of the events with an electron found in the 
LUMI at $\rts\simeq 91\GeV$ or
in the VSAT at $\rts\simeq 183\GeV$ 
is shown in Figure \ref{fig:malltags}. 
The number of events with a tag in the LUMI at $\rts\simeq 183 \GeV$ 
is too low to contribute to the measurement.
At $\rts\simeq 91\GeV$, the LUMI  
covers the $\qsq$ range from 1.3 $\GeV^{2}$ to 8.5 $\GeV^{2}$.
The VSAT covers at $\rts\simeq 183\GeV$ the $\qsq$ range between
0.2 \GeV$^{2}$ and 0.8 \GeV$^{2}$.
The solid line in Figure \ref{fig:malltags}
represents a fit with an exponential 
for the background and a Gaussian for the signal, 
with the position and width of the $\etac$ Gaussian fixed to the Monte Carlo
values.
The area of the Gaussian is $8.3^{+5.5}_{-4.9}$ events.

We assume that the shape of the $\etac$ form factor is described by the Vector 
Meson Dominance model form factor given in Equation (\ref{eq:ff}).
We compare three hypotheses for the pole mass $\Lambda$: 
a $\rho$-pole, which is found to 
be a good description of the $\qsq$ dependence of the 
$\pi^{0}$, $\eta$, and $\eta'$ cross sections \cite{ffetap}, 
a $\jpsi$-pole, which is 
predicted to be a good approximation of 
the $\qsq$ dependence of the $\etac$ cross section \cite{feldmann,schuler}, 
and an infinite pole or flat form factor, \ie\ the $\gaga\ra\etac$
cross section has no $\qsq$ dependence.
Since our $\etac$ Monte Carlo has been generated with a $\jpsi$-pole, the 
Monte Carlo events are reweighted to simulate a $\rho$-pole or flat 
form factor.

From Equations (\ref{eq:sigmatot}) and (\ref{eq:sigmagg}) 
it can be seen that the 
cross section is proportional to the product of the two-photon width 
and the form factor.
In order to measure the form factor in a $\qsq$-interval $\Delta\qsq$,
the ratio $\sigma_{\rm data}(\Delta\qsq)/\Ggg$ has to be determined.
The two-photon width $\Ggg$ has already been measured using 
untagged events, and the cross section $\sigma_{\rm data}(\Delta\qsq)$
can be obtained using the tagged events.
Note that in the ratio $\sigma_{\rm data}(\Delta\qsq)/\Ggg$
almost all systematic errors cancel, 
in particular the error due to 
the uncertainty in BR($\jpsi \ra \etac\gamma$).

The two cross sections $\sigma_{\rm data}(\Delta\qsq)$, 
for events with a 
tag in the LUMI and for events with a tag in the VSAT, are 
obtained using an unbinned likelihood fit.
The fit is similar to the one used to obtain the two-photon width, 
with the number of tagged signal events in each spectrum, $S_{i}$, equal to
$\epsilon_i(\Delta\qsq) \: \mathcal{L} \: {\rm BR}_i \: 
\sigma_{\rm data}(\Delta\qsq)$. 
The efficiencies per $\qsq$ interval, $\epsilon_{i}(\Delta\qsq)$, 
have an uncertainty 
due to the choice of the form factor in the Monte Carlo less than 3\%.
The cross sections correspond to 7.7$\pm$3.0 events with a tag in the LUMI 
and 2.3$\pm$2.3 events with a tag in the VSAT.

The $\sigma_{\rm data}(\Delta\qsq)/\Ggg$ ratios
are given in Table \ref{table:formfactor}, together
with the Monte Carlo predictions for the three form factor hypotheses.
A $\chi^{2}$ representing how well the measured
$\sigma_{\rm data}(\Delta\qsq)/\Ggg$
correspond to the Monte Carlo predictions
is calculated  and the corresponding 
probabilities are given in the last column of Table \ref{table:formfactor}
for each form factor hypothesis.
The changes in the $\chi^{2}$ probabilities due to the variation of 
$\Ggg$ within its measured error ($\pm$1.7 \keV) are given in parentheses 
in the last column.
The $\jpsi$-pole form factor and the flat form factor are clearly
favoured over the $\rho$-pole form factor.

Since $\sigma(\Delta\qsq)$ is proportional to
$F^{2}(\qsq)$, the latter can be obtained by dividing the
measured cross section (normalised to the measured two-photon width) 
by the Monte Carlo cross section with a flat form 
factor, for which $F^{2}=1$,
\beq
F^{2}(\qsq) = \frac{\sigma_{\rm data}(\Delta\qsq)}
              {\sigma_{\rm flat \, MC}(\Delta\qsq)}.
\eeq
The $F^{2}(\qsq)$ measured in the two $\qsq$ intervals is shown in 
Figure \ref{fig:ffjpsi}, together with three curves corresponding to
the theoretical predictions
for a $\rho$-pole, $\jpsi$-pole and flat form factor.
For a $\jpsi$-pole Monte Carlo, the average $\qsq$ for events 
with a tag in the LUMI is 3.08 \GeV$^{2}$, and for events with a tag in the 
VSAT it is 0.42 \GeV$^{2}$.
If the data are fitted with Equation (\ref{eq:ff}) 
with the pole mass $\Lambda$ left free, we obtain 
$\Lambda = 5.3^{+\infty}_{-2.5} \GeV$. The 95\% confidence level lower 
limit on the pole mass is 1.6 \GeV.

\section*{Conclusions}

The charmonium resonance $\etac$ is observed through
the reconstruction of nine different decay modes
at $\rts = 91 \GeV$ and at $\rts = 183 \GeV$.
The two-photon width is determined to be    
$\Ggg(\etac) = 6.9\pm 1.7 \stat \pm 0.8 \sys \pm 2.0 ({\rm BR})$ \keV,
in agreement with earlier measurements.
This corresponds to 76$\pm$19 signal events in total.
A $\chiczero$ signal of 29$\pm$11 events is also observed, but 
$\Ggg(\chiczero)$ cannot be evaluated since its branching ratios are unknown.
We find upper limits $\Ggg(\chiczero)<5.5 \keV$, $\Ggg(\chictwo)<1.4 \keV$,
and $\Ggg(\etac')<2.0 \keV$.
Using tagged $\etac$ events
we establish that the $\qsq$ dependence of the
cross section is better described by
a $\jpsi$-mass pole in the Vector Meson Dominance 
form factor than by a $\rho$-mass pole.

%
%
\section*{Acknowledgements}

We wish to 
express our gratitude to the CERN accelerator division for
the excellent performance of the LEP machine. 
We acknowledge the contributions of the engineers 
and technicians who have participated in the construction 
and maintenance of this experiment.

%
%
\newpage
\section*{Author List}
\input namelist175.tex
\newpage
%
%
\bibliographystyle{l3stylem}

\newpage


\begin{table}
  \begin{center}
    \begin{tabular}{|l|c|c|c|c|c|} \hline
      Decay mode & \multicolumn{1}{|c|}{BR(\%)} 
         & \multicolumn{1}{c|}{$\epsilon_{91}$ (\%)} 
         & \multicolumn{1}{c|}{$\epsilon_{183}$ (\%)}
         & \multicolumn{1}{c|}{$\Ggg^{91} (\keV)$ }
         & \multicolumn{1}{c|}{$\Ggg^{183} (\keV)$ } \\ 
      \hline \hline 
$\etac \ra \pipi\pipi$        & 1.2 $\pm$ 0.4   & 4.9 & 4.1
  & $<28$ & $<36$ \\ 
$\etac \ra \kk\pipi$                  & 2.0 $\pm$ 0.7   & 5.2 & 4.5
  & $10\pm 8$ & $16\pm 10$ \\ 
$\etac \ra \k0kpi$                    & 1.3 $\pm$ 0.4   & 5.1 & 5.0
  & $5.4^{+3.9}_{-3.4}$ & $5.5^{+4.7}_{-3.9}$ \\ 
$\etac \ra \pipi \eta (\gaga)$        & 1.3 $\pm$ 0.5   & 3.3 & 3.9 
  & $7.4^{+4.7}_{-4.4}$ & $<12$ \\
$\etac \ra \kk\pi^{0}$                & 0.9 $\pm$ 0.3   & 2.6 & 3.3
  & $21^{+12}_{-11}$ & $<34$ \\ 
$\etac \ra \rhrh$                     & 1.7 $\pm$ 0.6   & 0.8 & 0.9
  & $<28$ & $24^{+21}_{-17}$ \\
$\etac \ra \pipi \eta (\pipi\pi^{0})$ & 0.8 $\pm$ 0.3 & 1.6 & 1.2
  & $<16$ & $15^{+18}_{-11}$ \\
$\etac \ra \pipi\eta' (\pipi\eta)$    & 0.5 $\pm$ 0.2 & 2.8 & 2.5
  & $7^{+7}_{-6}$ & $<32$ \\
$\etac \ra \pipi\eta' (\rho \gamma)$  & 0.8 $\pm$ 0.3 & 3.0 & 2.5
  & $16\pm 10$ & $<29$ \\ 
        \hline
    \end{tabular}
    \caption{%
The branching ratio BR (derived from Reference \protect\cite{PDG98}), 
the efficiencies $\epsilon_{\rts}$, and the two-photon widths
as obtained with an unbinned likelihood fit,
at $\rts\simeq 91 \GeV$ and $\rts\simeq 183 \GeV$
for the different $\etac$ decay modes considered in the analysis.
All branching ratios contain a common error of 28\% due to the uncertainty 
in BR($\jpsi\ra\gamma\etac$).
    \label{table:result}}
  \end{center}
\end{table}

\begin{table}
  \begin{center}
    \begin{tabular}{|l|c|} \hline
      Experiment &\multicolumn{1}{c|}{$\Ggg(\etac)$ (\keV)}\\ 
      \hline \hline
      PLUTO & 28 $\pm$ 15 \\
      TPC/2$\gamma$ & $6.4^{+5.0}_{-3.4}$ \\
      CLEO & $5.9^{+2.1}_{-1.8}\pm 1.9$  \\
      L3 & 8.0 $\pm$ 2.4 $\pm$ 2.3   \\  
      ARGUS & 11.3 $\pm$ 4.2 \\
      E760 & $6.7^{+2.4}_{-1.7} \pm 2.3$ \\
                                                            \hline 
      L3 (this analysis) & $6.9 \pm 1.7 \pm 0.8 \pm 2.0$ \\ 
            \hline
    \end{tabular}
    \caption{Summary of the published measurements of $\Ggg(\etac)$ 
              \protect\cite{others,l3etac}. 
    \label{table:comparison}}
  \end{center}
\end{table}

\begin{table}
 \begin{center}
  \begin{tabular}{|l|c|c|c|} \hline
    & \multicolumn{1}{c|}{$\sigma(\Delta Q^{2})/\Ggg$ (pb/keV)} 
    & \multicolumn{1}{c|}{$\sigma(\Delta Q^{2})/\Ggg$ (pb/keV)} 
    & \multicolumn{1}{c|} {probability (\%)} \\ 
    & LUMI & VSAT & \\
\hline \hline
    Data & 1.3$\pm$0.5 & 5$\pm$5 &  \\ \hline
    MC $\rho$-pole & 0.051 & 0.81 & 4.3 (4.0 - 4.7)\\
    MC $\jpsi$-pole & 0.85 & 2.1 & 58 (34 - 81)\\
    MC flat & 1.6 & 2.3 & 70 (82 - 30)\\ \hline
  \end{tabular}
  \caption{Results for the tagged $\etac$ cross section 
$\sigma_{\rm data} (\Delta Q^{2})/\Ggg(\etac)$
for events with a tag in 
the LUMI at $\rts\simeq 91 \GeV$ and with a tag in the 
VSAT at $\rts\simeq 183 \GeV$, compared to the Monte Carlo 
predictions for a $\rho$-pole
form factor, a $\jpsi$-pole form factor, and a flat form factor.
In the last column the $\chi^{2}$ probability that the data and the 
Monte Carlo are compatible is given for each form factor hypothesis.
Between parentheses, the $\chi^{2}$ probabilities if $\Ggg$ is varied 
from 5.2 \keV\ to 8.6 \keV\ are given.
  \label{table:formfactor}}
 \end{center}
\end{table}


\begin{figure}[htbp]
  \begin{center}
    \includegraphics[width=0.98\textwidth]{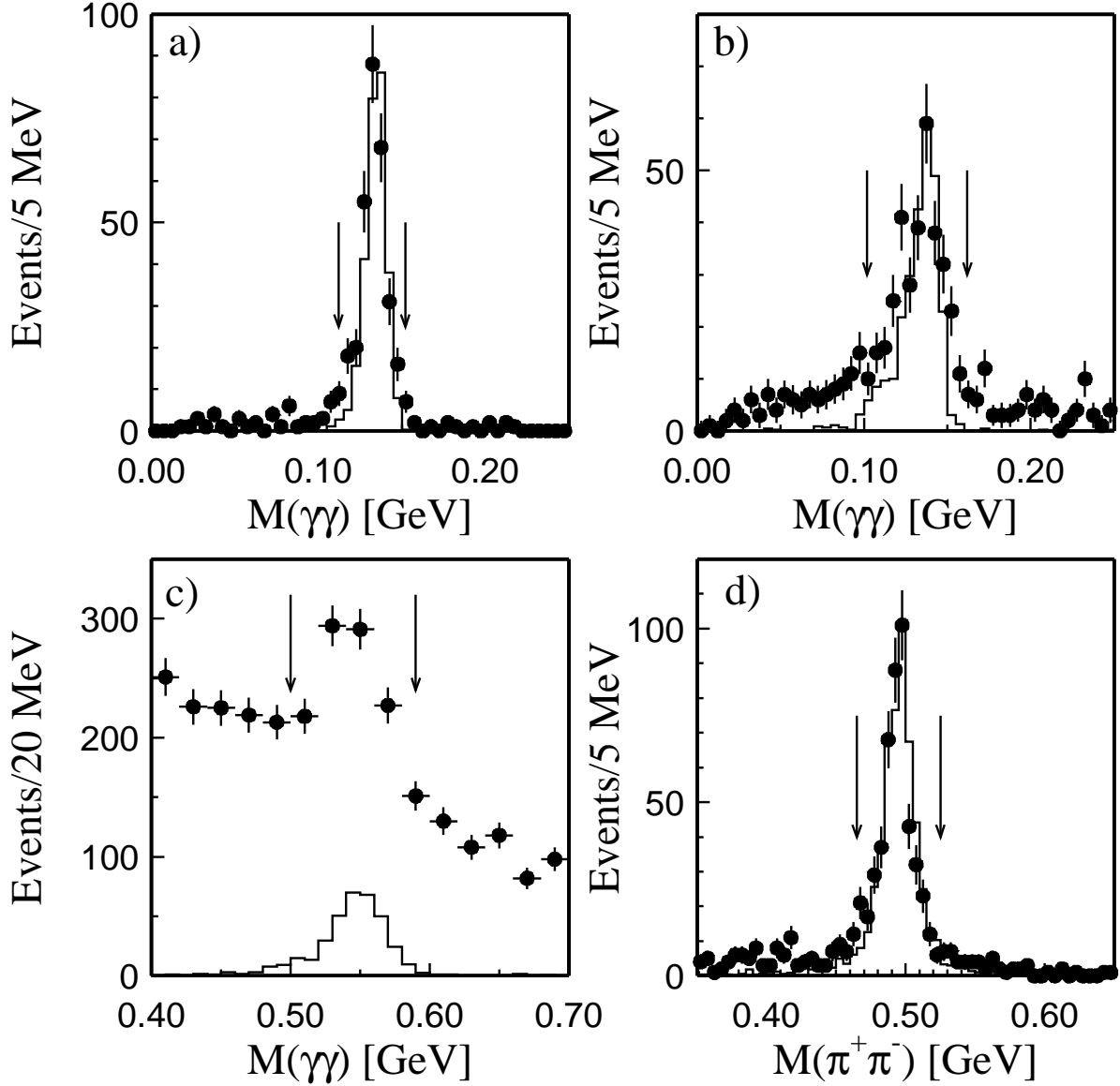}
  \end{center}
  \caption{a) The invariant mass of two photons in the $\kk\pi^{0}$ sample, 
    with both photons in the barrel ($42^{\circ}<\theta<138^{\circ}$)
    after all other cuts have been performed.  The events between the 
   arrows are accepted. b) The same, with at least 
 one of the photons in the endcaps ($12^{\circ}<\theta<38^{\circ}$ and 
 $142^{\circ}<\theta<168^{\circ}$). c) The invariant mass of two photons 
 in the $\pipi\eta$ sample, after all other cuts have been performed.
 d) The invariant mass of the two tracks associated with a secondary 
 vertex in the $\k0kpi$ sample, 
 selected with the cuts described in the text, after all other 
 cuts have been made.
In all plots the points with error bars represent the data, while 
the histogram represents the $\etac$ Monte Carlo. 
The normalisation of the Monte Carlo is arbitrary.}
  \label{fig:mcuts}
\end{figure}

\begin{figure}[htbp]
  \begin{center}
    \includegraphics[width=\textwidth]{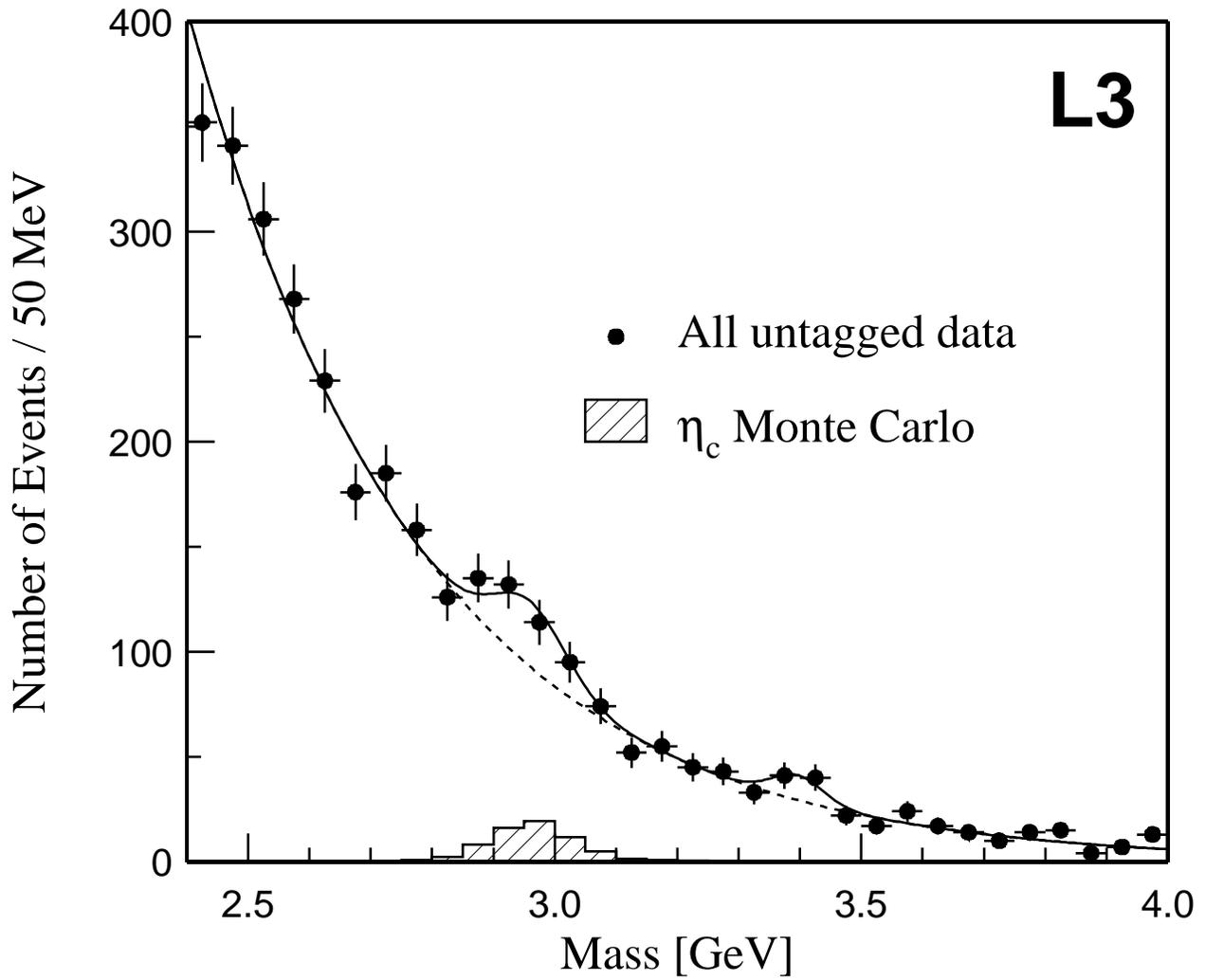}
  \end{center}
  \caption{The mass distribution of all selected untagged events.
 The hatched histogram is the arbitrarily 
scaled $\etac$ Monte Carlo. The solid lines represent a fit to the data
with an exponential background and Gaussians for the $\etac$ and $\chiczero$ 
signals. }
  \label{fig:m1etac}
\end{figure}

\begin{figure}[htbp]
  \begin{center}
    \includegraphics[width=\figwidth]{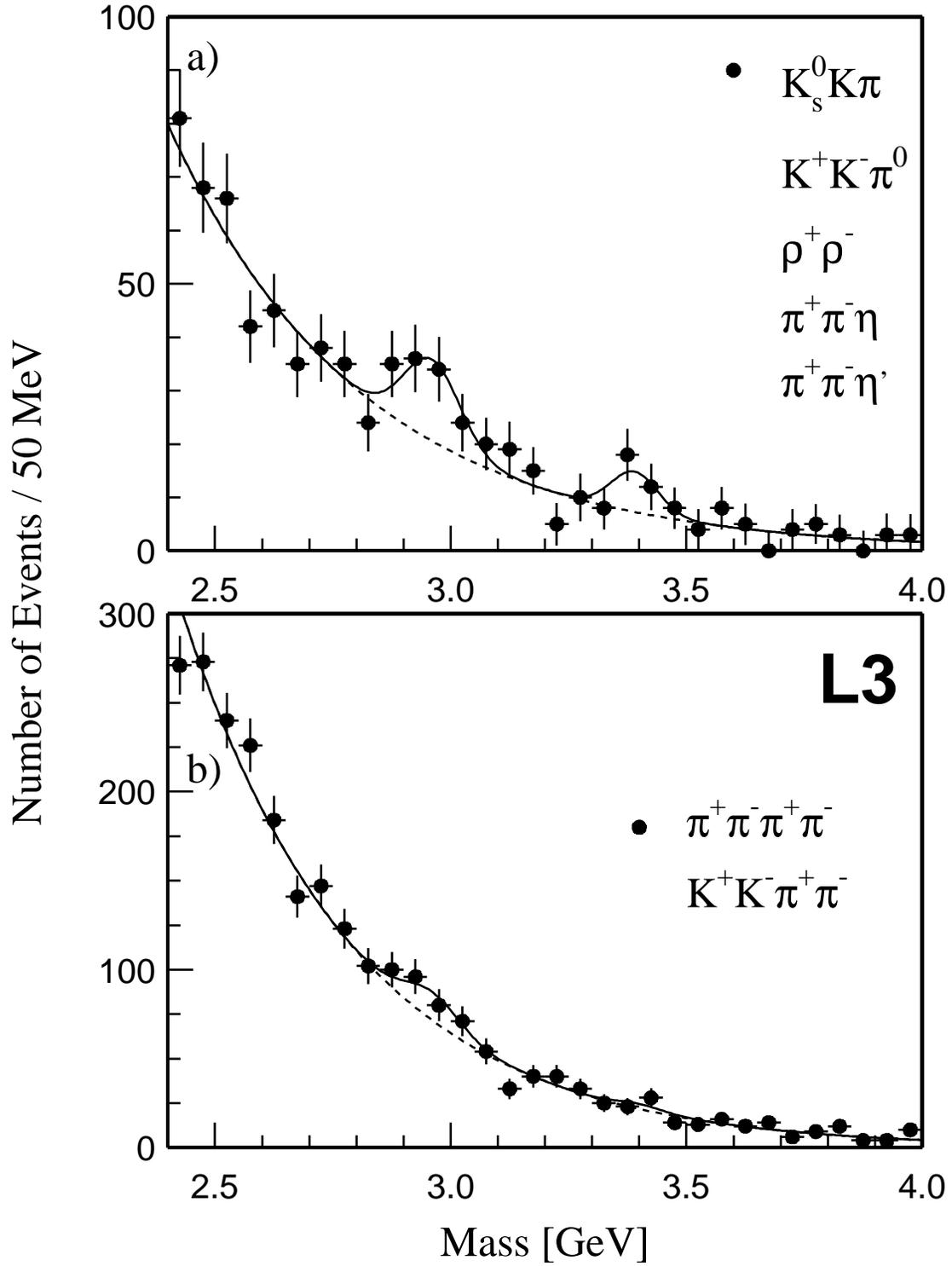}
  \end{center}
  \caption{The mass distribution of the selected untagged events a) 
for the decay modes $\k0kpi$, $\kk\pi^{0}$, $\rhrh$, $\pipi\eta$, 
and $\pipi\eta'$, b) for the decay modes $\pipi\pipi$ and $\kk\pipi$.
 The solid lines represent fits to the data
with an exponential background and Gaussians for the 
$\etac$ and $\chiczero$ signals. }
  \label{fig:m2etac}
\end{figure}

\begin{figure}[htbp]
  \begin{center}
    \includegraphics[height=0.5\textheight]{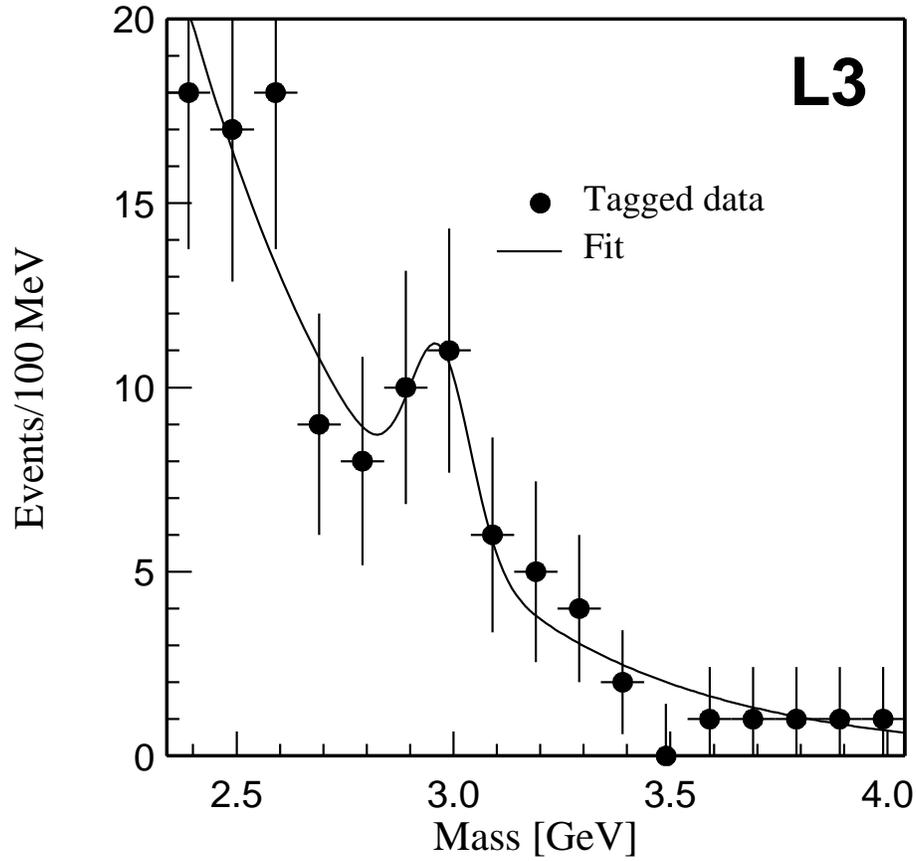}
  \end{center}
  \caption{The mass distribution for selected events with a tag in the 
LUMI at $\rts\simeq 91 \GeV$, or with a tag in the 
VSAT at $\rts \simeq 183 \GeV$. The line is a fit with an exponential 
background and a Gaussian for the signal, with the mass and the width of 
the Gaussian fixed to the $\etac$ Monte Carlo prediction.}
  \label{fig:malltags}
\end{figure}

\begin{figure}[htbp]
  \begin{center}
    \includegraphics[height=0.6\textheight]{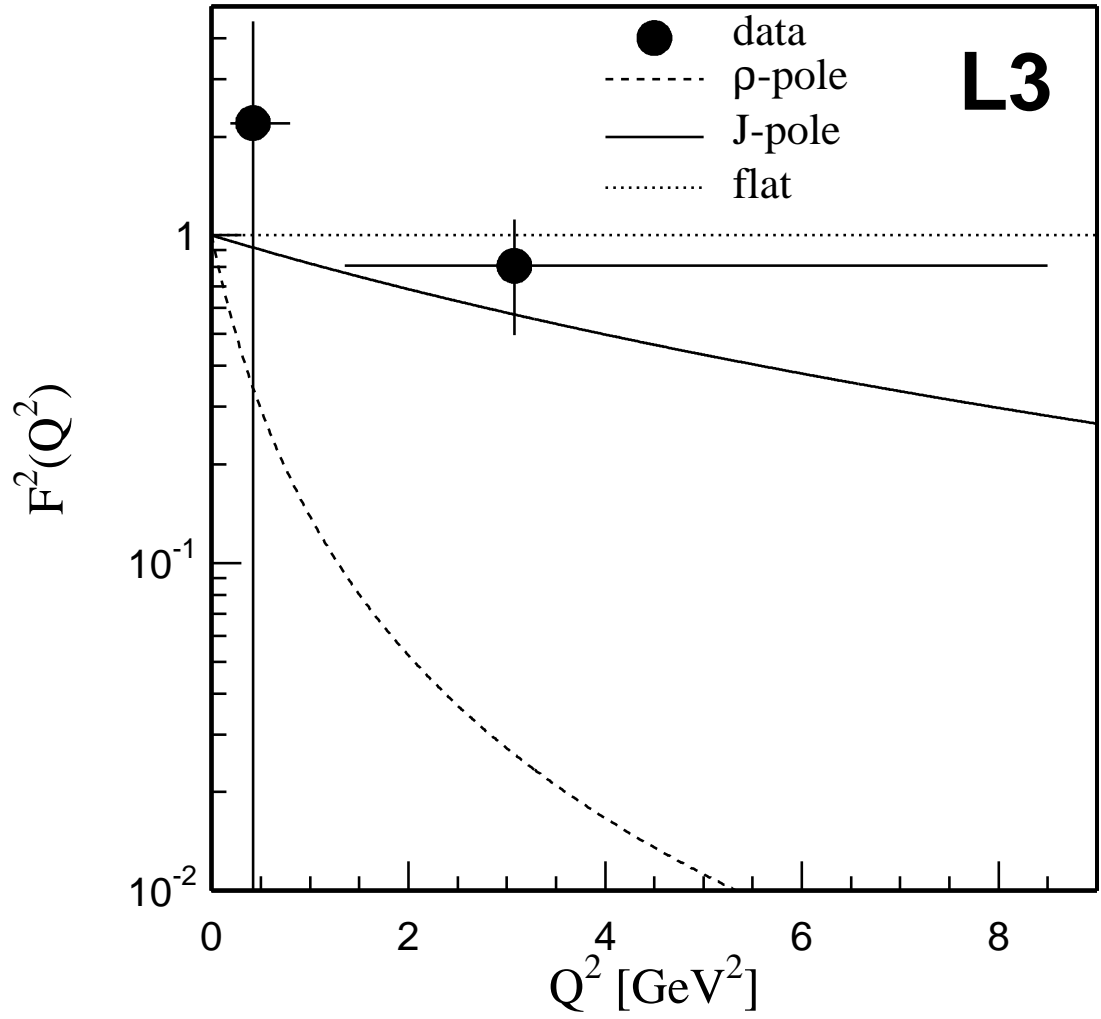}
  \end{center}
  \caption{The form factor $F^{2}(\qsq)$ \vs\ $\qsq$. The points represent 
 the tagged data, while the three curves describe the form factor shape for 
 a $\rho$-pole, $\jpsi$-pole or flat Monte Carlo.}
  \label{fig:ffjpsi}
\end{figure}

\end{document}

%% file: namelist175.tex
\typeout{   }     
\typeout{Using author list for paper 175 -?}
\typeout{$Modified: Tue May 18 08:38:13 1999 by clare $}
\typeout{!!!!  This should only be used with document option a4p!!!!}
\typeout{   }
%
%
%
%
%
%

\newcount\tutecount  \tutecount=0
\def\tutenum#1{\global\advance\tutecount by 1 \xdef#1{\the\tutecount}}
\def\tute#1{$^{#1}$}
\tutenum\aachen            
\tutenum\nikhef            
\tutenum\mich              
\tutenum\lapp              
\tutenum\basel             
\tutenum\lsu               
\tutenum\beijing           
\tutenum\berlin            
\tutenum\bologna           
\tutenum\tata              
\tutenum\ne                
\tutenum\bucharest         
\tutenum\budapest          
\tutenum\mit               
\tutenum\florence          
\tutenum\cern              
\tutenum\wl                
\tutenum\geneva            
\tutenum\hefei             
\tutenum\seft              
\tutenum\lausanne          
\tutenum\lecce             
\tutenum\lyon              
\tutenum\madrid            
\tutenum\milan             
\tutenum\moscow            
\tutenum\naples            
\tutenum\cyprus            
\tutenum\nymegen           
\tutenum\caltech           
\tutenum\perugia           
\tutenum\cmu               
\tutenum\prince            
\tutenum\rome              
\tutenum\peters            
\tutenum\salerno           
\tutenum\ucsd              
\tutenum\santiago          
\tutenum\sofia             
\tutenum\korea             
\tutenum\alabama           
\tutenum\utrecht           
\tutenum\purdue            
\tutenum\psinst            
\tutenum\zeuthen           
\tutenum\eth               
\tutenum\hamburg           
\tutenum\taiwan            
\tutenum\tsinghua          
{
\parskip=0pt
\noindent
{\bf The L3 Collaboration:}
\ifx\selectfont\undefined
 \baselineskip=10.8pt
 \baselineskip\baselinestretch\baselineskip
 \normalbaselineskip\baselineskip
 \ixpt
\else
 \fontsize{9}{10.8pt}\selectfont
\fi
\medskip
\tolerance=10000
\hbadness=5000
\raggedright
\hsize=162truemm\hoffset=0mm
\def\r{\rlap,}
\noindent

M.Acciarri\r\tute\milan\
P.Achard\r\tute\geneva\ 
O.Adriani\r\tute{\florence}\ 
M.Aguilar-Benitez\r\tute\madrid\ 
J.Alcaraz\r\tute\madrid\ 
G.Alemanni\r\tute\lausanne\
J.Allaby\r\tute\cern\
A.Aloisio\r\tute\naples\ 
M.G.Alviggi\r\tute\naples\
G.Ambrosi\r\tute\geneva\
% ...
\newpage
\begin{list}{A}{\itemsep=0pt plus 0pt minus 0pt\parsep=0pt plus 0pt minus 0pt
                \topsep=0pt plus 0pt minus 0pt}
\item[\aachen]
 I. Physikalisches Institut, RWTH, D-52056 Aachen, FRG$^{\S}$\\
 III. Physikalisches Institut, RWTH, D-52056 Aachen, FRG$^{\S}$
\item[\nikhef] National Institute for High Energy Physics, NIKHEF, 
     and University of Amsterdam, NL-1009 DB Amsterdam, The Netherlands
\item[\mich] University of Michigan, Ann Arbor, MI 48109, USA
\item[\lapp] Laboratoire d'Annecy-le-Vieux de Physique des Particules, 
     LAPP,IN2P3-CNRS, BP 110, F-74941 Annecy-le-Vieux CEDEX, France
\item[\basel] Institute of Physics, University of Basel, CH-4056 Basel,
     Switzerland
\item[\lsu] Louisiana State University, Baton Rouge, LA 70803, USA
\item[\beijing] Institute of High Energy Physics, IHEP, 
  100039 Beijing, China$^{\triangle}$ 
\item[\berlin] Humboldt University, D-10099 Berlin, FRG$^{\S}$
\item[\bologna] University of Bologna and INFN-Sezione di Bologna, 
     I-40126 Bologna, Italy
\item[\tata] Tata Institute of Fundamental Research, Bombay 400 005, India
\item[\ne] Northeastern University, Boston, MA 02115, USA
\item[\bucharest] Institute of Atomic Physics and University of Bucharest,
     R-76900 Bucharest, Romania
\item[\budapest] Central Research Institute for Physics of the 
     Hungarian Academy of Sciences, H-1525 Budapest 114, Hungary$^{\ddag}$
\item[\mit] Massachusetts Institute of Technology, Cambridge, MA 02139, USA
\item[\florence] INFN Sezione di Firenze and University of Florence, 
     I-50125 Florence, Italy
\item[\cern] European Laboratory for Particle Physics, CERN, 
     CH-1211 Geneva 23, Switzerland
\item[\wl] World Laboratory, FBLJA  Project, CH-1211 Geneva 23, Switzerland
\item[\geneva] University of Geneva, CH-1211 Geneva 4, Switzerland
\item[\hefei] Chinese University of Science and Technology, USTC,
      Hefei, Anhui 230 029, China$^{\triangle}$
\item[\seft] SEFT, Research Institute for High Energy Physics, P.O. Box 9,
      SF-00014 Helsinki, Finland
\item[\lausanne] University of Lausanne, CH-1015 Lausanne, Switzerland
\item[\lecce] INFN-Sezione di Lecce and Universit\'a Degli Studi di Lecce,
     I-73100 Lecce, Italy
\item[\lyon] Institut de Physique Nucl\'eaire de Lyon, 
     IN2P3-CNRS,Universit\'e Claude Bernard, 
     F-69622 Villeurbanne, France
\item[\madrid] Centro de Investigaciones Energ{\'e}ticas, 
     Medioambientales y Tecnolog{\'\i}cas, CIEMAT, E-28040 Madrid,
     Spain${\flat}$ 
\item[\milan] INFN-Sezione di Milano, I-20133 Milan, Italy
\item[\moscow] Institute of Theoretical and Experimental Physics, ITEP, 
     Moscow, Russia
\item[\naples] INFN-Sezione di Napoli and University of Naples, 
     I-80125 Naples, Italy
\item[\cyprus] Department of Natural Sciences, University of Cyprus,
     Nicosia, Cyprus
\item[\nymegen] University of Nijmegen and NIKHEF, 
     NL-6525 ED Nijmegen, The Netherlands
\item[\caltech] California Institute of Technology, Pasadena, CA 91125, USA
\item[\perugia] INFN-Sezione di Perugia and Universit\'a Degli 
     Studi di Perugia, I-06100 Perugia, Italy   
\item[\cmu] Carnegie Mellon University, Pittsburgh, PA 15213, USA
\item[\prince] Princeton University, Princeton, NJ 08544, USA
\item[\rome] INFN-Sezione di Roma and University of Rome, ``La Sapienza",
     I-00185 Rome, Italy
\item[\peters] Nuclear Physics Institute, St. Petersburg, Russia
\item[\salerno] University and INFN, Salerno, I-84100 Salerno, Italy
\item[\ucsd] University of California, San Diego, CA 92093, USA
\item[\santiago] Dept. de Fisica de Particulas Elementales, Univ. de Santiago,
     E-15706 Santiago de Compostela, Spain
\item[\sofia] Bulgarian Academy of Sciences, Central Lab.~of 
     Mechatronics and Instrumentation, BU-1113 Sofia, Bulgaria
\item[\korea] Center for High Energy Physics, Adv.~Inst.~of Sciences
     and Technology, 305-701 Taejon,~Republic~of~{Korea}
\item[\alabama] University of Alabama, Tuscaloosa, AL 35486, USA
\item[\utrecht] Utrecht University and NIKHEF, NL-3584 CB Utrecht, 
     The Netherlands
\item[\purdue] Purdue University, West Lafayette, IN 47907, USA
\item[\psinst] Paul Scherrer Institut, PSI, CH-5232 Villigen, Switzerland
\item[\zeuthen] DESY-Institut f\"ur Hochenergiephysik, D-15738 Zeuthen, 
     FRG
\item[\eth] Eidgen\"ossische Technische Hochschule, ETH Z\"urich,
     CH-8093 Z\"urich, Switzerland
\item[\hamburg] University of Hamburg, D-22761 Hamburg, FRG
\item[\taiwan] National Central University, Chung-Li, Taiwan, China
\item[\tsinghua] Department of Physics, National Tsing Hua University,
      Taiwan, China
\item[\S]  Supported by the German Bundesministerium 
        f\"ur Bildung, Wissenschaft, Forschung und Technologie
\item[\ddag] Supported by the Hungarian OTKA fund under contract
numbers T019181, F023259 and T024011.
\item[\P] Also supported by the Hungarian OTKA fund under contract
  numbers T22238 and T026178.
\item[$\flat$] Supported also by the Comisi\'on Interministerial de Ciencia y 
        Tecnolog{\'\i}a.
\item[$\sharp$] Also supported by CONICET and Universidad Nacional de La Plata,
        CC 67, 1900 La Plata, Argentina.
\item[$\natural$] Supported by Deutscher Akademischer Austauschdienst.
\item[$\diamondsuit$] Also supported by Panjab University, Chandigarh-160014, 
        India.
\item[$\triangle$] Supported by the National Natural Science
  Foundation of China.
\end{list}
}
\vfill


